\documentclass[12pt,a4paper]{article}
\usepackage{amssymb,amsfonts,amsmath}

\title{External observer reflections on QBism}

\author{Andrei Khrennikov\\
International Center for Mathematical Modelling \\
in Physics and Cognitive Sciences\\
Linnaeus University,
V\"axj\"o, SE-351 95, Sweden\\
Andrei.Khrennikov@lnu.se}

\date{}
 
\begin{document}

\maketitle

\begin{abstract} In this short review I present my personal reflections on QBism. 
I have no intrinsic sympathy neither to QBism  nor to subjective interpretation of probability 
in general. However, I  have been following development of QBism from its very beginning,
observing its evolution and success, sometimes with big surprise.  
Therefore my reflections on QBism can be treated as ``external observer'' reflections. 
I hope that my view on this interpretation of quantum mechanics (QM) has some 
degree of objectivity. It may be useful for researchers who are interested in quantum foundations,
but do not belong to the QBism-community, because I tried to analyze essentials of QBism critically
(i.e., not just emphasizing its advantages, as in a typical QBist publication).  
QBists, too, may be interested in comments of an external observer who monitored 
development of this approach to QM during the last 16 years. The second part of the paper is devoted 
to interpretations of probability, objective versus subjective, and views of Kolmogorov, von Mises, and 
 de Finetti. Finally, de Finetti's approach to methodology of science is presented and compared 
with QBism.  
\end{abstract}

\section{Introduction}

As is well known, QM which manifests huge success in the mathematical representation of the basic problems of physics of the micro-world, suffers heavily of diversity 
of interpretations of   this mathematical representation. QBism is one of the most recent attempts to provide a consistent interpretation of QM, free of all possible mysteries 
and paradoxes.   In this short review I present my personal reflections on QBism. The papers starts with a historical remark on the first years of QBism. It continues with brief 
representation of its essentials, mainly referring to and citing its creators. Then the main postulates of QBism are  critically analyzed. 
We emphasize that the cornerstone of QBism is the interpretation of QM as a machine for update of probabilities based on
 a modification of the classical formula of total probability  (FTP). QBism version of FTP is derived with the aid of  symmetric informationally complete  positive operator valued 
measures (SIC-POVMs).\footnote{ From this viewpoint, i.e., QM as a probability update machine, QBism is similar to the V\"axj\"o interpretation of QM. The latter is a realist contextual statistical 
interpretation, so ideologically it is opposite to QBism. However,  the probability update basis makes it close to QBism. At the same time, even this closeness is only formal, since
the QBism version of generalized  FTP differs crucially from the V\"axj\"o version.    } 

The rest of the paper is devoted to 
interpretations of probability. Here diversity is not so huge as in QM, but the gap between two main interpretations, objective and subjective, is no smaller than the gap between
two basic trends in interpretations of QM, realist (in the spirit of Einstein) and non-realist (in the spirit of Copenhagen). We present the original Kolmogorov interpretation of probability \cite{K, K1, K2}
(which is not so well known, even in the probability community) and compare it with the genuine frequency interpretation of von Mises \cite{[169], [170], [171]} 
and subjective interpretation of de Finetti \cite{Finetti}, \cite{Finetti1}.

Then, following de Finetti,   we point out that consistent appealing to the subjective interpretation of probability should lead to reconsideration of the objective treatment of  
the scientific methodology. Finally we point that de Finetti was even more revolutionary than QBists, because his subjective treatment of scientific method was not restricted 
to ``special quantum world''. In some way QBists made one good turn but refrain from another, they revolutionary declared the private agent (user) perspective to knowledge about ``quantum 
world'', but they were not brave enough to follow de Finetty completely, i.e., to declare the private agent perspective for knowledge about classical world as well.
The latter is an exciting project still waiting its realization.

\section{QBism childhood  in V\"axj\"o}

In 2001 QBism \index{QBism} was strongly represented   at the second V\"axj\"o conference on quantum foundations, ``Quantum Theory: Reconsideration of Foundations'' (QTFT2001), 
June 17-21, 2001. We (organizers and participants of this conference)  strongly believed that the quantum information revolution would soon lead to great 
foundational revolution.  Unfortunately, dreams did not come true.  Nevertheless,  the energy of the quantum information revolution was transformed in a series of stormy 
debates during the series of the V\"axj\"o conferences, 2000-2015.  Although these debates did not 
lead to a complete resolution of the basic problems of quantum mechanics, they clarified some of these problems, especially the problem of the interpretation of a quantum state. 
QBism was definitely one of the main foundational outputs of the quantum information revolution.
\footnote{Besides QBism, we can mention the {\it V\"axj\"o interpretation} of QM (statistical realist and contextual) \cite{V1}, \cite{V2} derivation of the QM-formalism 
from simple operational principles, D' Ariano \cite{D1}, \cite{D2} and  Chiribella et al. \cite{Ch1}, \cite{Ch2}
  (first time this project was also announced in V\"axj\"o), and the {\it statistical Copenhagen interpretation} 
(statistical non-realist) which final formulation  was  presented at the V\"axj\"o-15 conference by A. Plotnitsky 
and based on his previous studies about the probabilistic structure of QM \cite{PL0}--\cite{PL3}.}

As the organizers of QTFT-2001, I and C. Fuchs both dreamed for creation of a consistent and clear interpretation of QM, free of mysteries and paradoxes. However, we went in two opposite 
directions. I followed Einstein and later, as the result of better understanding of Bohr's writings (and especially comments of A. Plotnitsky on them), tried to unify Einstein's realist 
statistical interpretation with Bohr's contextual interpretation by filtering out Bohr's nonrealist attitude, see  \cite{V1, V2, KHR_CONT} for so called 
{\it V\"axj\"o interpretation} \index{V\"axj\"o interpretation} 
of QM.  Both Einstein and Bohr (as well as, e.g., von Neumann) used the statistical (ensemble) interpretation of quantum probabilities. Therefore the V\"axj\"o interpretation is based 
on this statistical interpretation. It also was  important that as student I was strongly influenced by A. N. Kolmogorov and  B. V. Gnedenko who always emphasized that probability 
is objective and statistical. Later, after PhD,  I discovered works of   von Mises \cite{[169], [170], [171]}. 
I really enjoyed this reading! 
 Von Mises' interpretation of probability differs from Kolmogorov's interpretation.
Nevertheless, Misesian probability is also objective and statistical . In any event, nobody of them (Einstein, Bohr, von Neumann, Kolmogorov, Gnedenko) and neither I would agree with 
De Finetti's slogan: {\it ``Probability does not exist!''} and with his subjective interpretation of probability.  
My views on interpretation of quantum states and probabilities were presented in \cite{V1}.  

C. Fuchs went in the opposite direction,
he (with support of C. M.  Caves, R. Schack, and D. Mermin) openly, loudly, and proudly declared \cite{Caves1}, \cite{Caves2}, \cite{Fuchs2a}- \cite{Fuchs6} that QM is only about knowledge (and here QBists are very close to the fathers of the 
Copenhagen interpretation, N. Bohr, and W. Heisenberg). But this widely supported viewpoint was completed with very strained and revolutionary declaration that this quantum knowledge 
has to be treated as {\it personal knowledge.}  Subjective interpretation of quantum probability matches perfectly such a private agent perspective of quantum theory.  
QBists emphasize the Bayesian probability update and decision making structure of the quantum probability calculus. 

From my viewpoint, the latter is one of the main contributions of QBism in clarification of quantum foundations. Independently this viewpoint was presented in the framework of the 
V\"axj\"o interpretation, section \cite{V1, V2, KHR_CONT}.

At the beginning I and C. Fuchs did not recognize this similarity, namely, interpretation of the calculus of quantum probabilities as a machinery for update of probabilities. 
And it is clear why: at that time I emphasized realism and objectivity and C. Fuchs privacy and subjectivity. And the probability update dimension was shadowed by these
philosophic issues.   This explains the appearance of  Fuch's anti-V\"axj\"o paper \cite{Fuchs2}. 

Nevertheless, intuitively I felt sympathy to QBism but roots of this sympathy were not clear for me.\footnote{Once Christopher Fuchs asked me: ``Why did you support QBism so strongly during 
the V\"axj\"o-series of conferences? QBism contradicts your own V\"axj\"o interpretation!'' In fact, I was not able to explain this even for myself.  I had a feeling that QBIsm can be useful. 
But how? and where? } At the  V\"axj\"o-15 conference  QBism was widely represented and celebrated its world-wide recognition. Nobel Prize Laureate T. H\"ansch 
presented the great lecture  about QBism as the only possible consistent foundational basis of quantum information theory.  
This lecture ignited the stormy debate and T. H\"ansch
and C. Fuchs were attacked by the realist opposition (leaded by L. Vaidman and A. Elitzur).      
 
\section{Quantum theory is about evaluation of  expectations for
the content of personal experience}
\label{QBINT}

In contrast to von Neumann, Fuchs proposed to interpret probability in the subjective way.
To present essentials of some theory, sometimes it is practical simply to cite works of its creators (this is definitely not 
the case of Bohr,  or von Neumann, or even myself). Here we cite C. Fuchs and R. Schack \cite{Fuchs5}, pp. 3-4:

{\small ``The fundamental primitive of QBism\index{QBism} is the concept of experience. According to QBism,
quantum mechanics is a theory that any agent can use to evaluate her expectations for
the content of her personal experience.

QBism adopts the personalist Bayesian probability theory pioneered by Ramsey \cite{[12a]} and de Finetti \cite{Finetti} 
and put in modern form by Savage \cite{[14a]} and Bernardo and Smith \cite{[15a]} among others. This means that QBism interprets all probabilities, in particular
those that occur in quantum mechanics, as an agent's personal, subjective degrees of
belief. This includes the case of certainty - even probabilities 0 or 1 are degrees of belief. ...

In QBism, a measurement is an action an agent takes to elicit an experience.
The measurement outcome is the experience so elicited. The measurement outcome
is thus personal to the agent who takes the measurement action. In this sense, quantum
mechanics, like probability theory, is a single user theory. A measurement does not reveal
a pre-existing value. Rather, the measurement outcome is created in the measurement
action.

According to QBism, quantum mechanics can be applied to any physical system.
QBism treats all physical systems in the same way, including atoms, beam splitters,
Stern-Gerlach magnets, preparation devices, measurement apparatuses, all the way to
living beings and other agents. In this, QBism differs crucially from various versions
of the Copenhagen interpretation. ...

An agent's beliefs and experiences are necessarily local to that agent. This implies
that the question of nonlocality simply does not arise in QBism.''} 

\medskip

We shall revisit the interpretational issues of QBism after the presentation of its basic 
probabilistic principle in the next session. 
 
\section{QBism as a probability update machinery}
\label{UP}

The previous section might create the impression that  the subjective interpretation of quantum 
probabilities is the key point of QBism. It might be that even its creators have the same 
picture of their theory.  For me, the essence of QBism is neither this very special interpretation of quantum 
probabilities nor the concrete agent perspective of QM. For me, the main ideological invention of 
C. Fuchs and R. Schack was treatment of the mathematical formalism of QM as a generalization 
of the classical Bayesian machinery of the probability update. This viewpoint clarifies the meaning of 
the basic rule of QM -- the Born rule as a complex Hilbert space representation of  
generalization of the {\it classical formula of total probability} 
(FTP).  

This dimension of QBism is identical to the probability update dimension of the V\"axj\"o 
interpretation \cite{V1, V2, KHR_CONT}.  There are a few differences. 

One is the difference in 
the interpretations of probability. However, nowadays I do not consider it as the crucial difference (in contrast to my 
first debates with C. Fuchs in 2001-2003, \cite{V1} and \cite{Fuchs2}). Really, in a long series of updates the subjective
and statistical viewpoints coincide. \footnote{
In the mentioned debates I also was strongly against the anti-realist attitude 
of QBism. However,  now this attitude  does not disturb me so much as 12 year ago. Either I started to understand
QBists views on the problem of realism better or QBists changed their views (or both). QBism needs not appeal to any subquantum 
model providing the ontic description of  quantum systems and processes, in particular, to hidden variables.
Nor is QBism concerned with struggle against such models. It seems that the personal position of 
C. Fuchs is similar to the position of N. Bohr: for quantum {\it physics,} 
it plays no role whether  finally one would be able to construct a realistic subquantum model or not.   
For the present state of development of quantum theory, this is the most reasonable position, cf. 
with Zeilinger's strong anti-realist attitude.  
Moreover, just recently (through a series of email exchanges) I understood better the position of C. Fuchs  
on the problem of non/realism. Surprisingly QBists (at least C. Fuchs) do not consider QBism as a non-realist 
interpretation of QM. } 

For me, the main difference between QBism and the V\"axj\"o interpretation 
is in the mathematics, not in physics or in philosophy.   Both 
the V\"axj\"o interpretation of QM \cite{[16]}, \cite{[17]}, \cite{KHR_CONT} 
and QBism \cite{Fuchs3}-\cite{Fuchs6}  {\it the Born rule is treated as generalization of classical formula of total probability}
 (FTP) in the language of linear operators. However, these interpretations are based on two totally different mathematical generalizations
of FTP - both matching the Born rule.
 
Starting with the Born rule,  QBists derived their special version of 
generalized FTP which is based on a very special class of the quantum probability updates, based on 
atomic instruments with SIC-POVMs, see appendix.

We now briefly present the QBism scheme for the probability update, namely, the representation of the Born rule as 
a generalization of FTP; here we again follow Fuchs and Schack \cite{Fuchs3}-\cite{Fuchs5}.\footnote{We remark that in 
coming considerations the interpretation of probabilities  does not play any role.
They can be subjective probabilities (as originally in QBism), but they also can be statistical, e.g., Kolmogorovian or 
Misesian, as well. }   

\medskip

    Quantum states are represented by density operators $\rho$ in a Hilbert space assumed
to be finite dimensional. A measurement (an action taken by the agent) is described by
a POVM $F=(F_j)$, where $j$ labels the potential outcomes experienced by the agent. The
agent's personalist probability\footnote{As was remarked, probability can be interpreted in other ways. The situation is similar to the classical probability update. 
De Finetti would treat this probability as subjective, but Kolmogorov, or Gnedenko, or von Mises as statistical.} 
$p(F_j)$ of experiencing outcome $j$  is given by the Born rule:
\begin{equation}
\label{BR2}
p(F_j) = \rm{Tr}  F_j  \rho.
\end{equation}
 Similar to the probabilities on the left-hand side of the Born rule, QBism regards the
operators $\rho$ and $F_j$ on the right-hand side as judgments made by the agent, representing
her personalist degrees of belief.

The Born rule as written in Eq. (\ref{BR2}) appears to connect probabilities on the left-hand
side of the equation with other kinds of mathematical objects - operators - on the right hand side.

QBists assume that the agent's reference measurement is an arbitrary \index{informationally complete}{\it informationally
complete POVM}, $E=(E_i),$  such that each $E_i$ is of rank 1, i.e., is proportional to a
one-dimensional projector. Such measurements exist for any finite Hilbert-space
dimension.  Furthermore, we assume that, if the agent carries out the measurement $E=(E_i)$
for an initial state $\rho,$ upon getting outcome $E_i$ he would update to the post-measurement
state 
$$
\rho_i= \frac{E_i \rho E_i}{\rm{Tr} E_i \rho E_i}.
$$ 
This is the assumption of atomicity of this quantum instrument. 
(This is a strong constraint on the class of instruments which are used in QBism. It would be interesting to analyze the possibility to proceed 
with arbitrary instruments.)

Because the reference measurement is informationally
complete, any state $\rho$ corresponds to a unique vector of probabilities 
$$
p(E_i) = \rm{Tr}  E_i \rho,
$$
and any POVM $F=(F_j)$ corresponds to a unique matrix of conditional probabilities
$$
p(F_j\vert E_i) = tr(F_j \rho_i) .
$$
The operators $\rho$ and $F_j$ on the right-hand side of the Born rule are thus
mathematically equivalent to sets of probabilities $p(E_i)$ and conditional probabilities
$p(F_j\vert E_i).$ (We remark that all these probabilities depend on the state $\rho,$ i.e., $p(E_i)\equiv p_\rho(E_i), p(F_j)\equiv p_\rho(F_j),
p(F_j\vert E_i)= p_\rho(F_j\vert E_i)).$  

Then Fuchs and Schack \cite{Fuchs3}-\cite{Fuchs5} stress that  POVMs as well as quantum states represent an agent's personal degrees of
belief.  However, this is not essential for the formal scheme of probability update. We can as well interpret  the probabilities 
$p(E_i)$ and $p(F_j\vert E_i)$ statistically.  The main point which was rightly emphasized by them is that the Born rule 
can be interpreted as one special form of transformation of probabilities: 
\begin{equation}
\label{QB1}
p(F_j) = f(p(E_i), p(F_j\vert E_i)).
\end{equation}
Comparing with the classical FTP which is in these notations written as
\begin{equation}
\label{QB2}
p(F_j) = \sum_ i p(E_i) p(F_j\vert E_i)
\end{equation}
they formulate the statement which I consider as the cornerstone of QBism: 

\medskip

{\it The Born rule is one of the forms of generalization of FTP.}

\medskip

For me, the main problem of  QBists is that they started with a SIC-POVM $E=(E_i).$ They say \cite{Fuchs3}: 
{\small ``The Born rule allows the agent to calculate her outcome probabilities
$p(F_j)$ in terms of her probabilities $p(E_i)$ and $p(F_j\vert E._i)$ defined with respect to a counterfactual
reference measurement.''} This reference to counterfactuals is really redundant. Why should SIC-POVM
measurement appear at all? As in the V\"axj\"o approach, one can start with an arbitrary 
POVM-measurement, say  $G=(G_i),$ to define probabilities $p(G_i)= \rm{Tr} \rho G_i$ 
providing information about the state $\rho.$

By taking into account such a possibility of generalization of QBist consideration\footnote{However, for QBists the above generalization - to start the probability update scheme 
with an arbitrary POVM  measurement $G=(G_j)$ and not with a SIC-POVM $E=(E_i)$  - seems to be unacceptable. 
They are really addicted on SIC-POVMs and on completeness of information gained at the first step, information about the state, even 
at the price of appearance of counterfactuals.}
I completely agree  with the following statement of  QBism \cite{Fuchs3}: 

{\small ``In QBism, the Born rule functions as a coherence requirement. Rather than setting
the probabilities $p(F_j),$ the Born rules relates them to those defining the state $\rho$ and
the POVM $F=(F_j)$. Just like the standard rules of probability theory, the Born rule is
normative: the agent ought to assign probabilities that satisfy the constraints imposed
by the Born rule.''}

The functional
relationship given by (\ref{QB1}) depends on details of the reference measurement. In the special
case that the reference measurement is a symmetric informationally complete POVM
(SIC) , (\ref{QB1}) takes the simple form:
\begin{equation}
\label{SIC}
p(F_j) = \rm{Tr} F_j \rho_i = \sum_i \Big( (d+1) p(E_i) - \frac{1}{d}\Big) p(F_j\vert E_i).
\end{equation}
This is a consequence of the complete information representation of a quantum state and the 
use of SIC-POVMs, see appendix.

QBists have conjectured that this form of the Born rule may be used as an axiom
in a derivation of quantum theory. The key question that remains is in identifying what minimal further
principles must be added to the equation (\ref{SIC}) for the project to be successful.

\medskip

{\bf Remark.} This program is identical to attempts to justify the V\"axj\"o interpretation  by deriving the 
complex Hilbert space formalism from the generalized FTP with an interference term \cite{INT0}, \cite{INT_KHR}, 
\cite{KHR_CONT}. 
However,  I  
tried to derive  the complex Hilbert space structure of QM  solely from the latter 
generalization of FTP, i.e., without additional axioms. This approach was successful only for 
dichotomous observables. Already the case of three-valued observables is very difficult mathematically.
Here only a partial succes was achieved by P. Nyman and I. Basieva \cite{Nyman0}, \cite{Nyman}. 
May be Fuchs and Schack are right that one has to 
find additional axioms leading to the complete derivation of the quantum formalism, 
as, e.g., was done by D' Ariano et al. \cite{D1}, \cite{D2}, \cite{Ch1}, \cite{Ch2}. However, proceeding with such additional 
axioms does not match completely the basic principle that the quantum formalism is just a special form of  
the probability update generalizing the classical Bayesian update. This principle 
is very attractive in both QBism and the V\"axj\"o interpretation (though they have different 
mathematical realizations).  Completing a generalized probability update scheme
by additional operational principles diminishes the value of this scheme as the {\it unique fundamental 
principle} lying in the ground of QM.  It shifts the line of research to more traditional operational 
approaches starting with  the pioneer contribution of Heisenberg  \cite{[7]}, then Mackey \cite{M7} and nowadays
D' Ariano et al. \cite{D1}, \cite{D2}, \cite{Ch1}, \cite{Ch2}.   In this paper we do not plan to analyze the operational axiomatics of D'Ariano et al..  
However, in previous derivations of QM from ``natural operational principles'' complex Hilber space was always 
encrypted, practically explicitly,   in one of the axioms, see, e.g., G. Mackey \cite{M7} for one of the first 
operational derivation  of the quantum formalism. There is still some hope that at least the V\"axj\"o scheme 
of the probability update, for the FTP with the interference term  \cite{INT0}, \cite{INT_KHR}, 
\cite{KHR_CONT}.  may lead to derivation of the complete formalism.
The difficulties which stopped my further studies in this direction were  of the mathematical origin -- too  complicated mathematical  
calculations. \index{D' Ariano principles} \index{Mackey principles}   

\section{Agents constrained by Born's rule}
\label{CST}

In 2001 the private user's experience viewpoint on quantum {\it physics}  made me really mad and this was the main reason 
for my anti-QBism attitude \cite{V1} (see Appendix A).\footnote{The strong anti-Copenhagen attitude in the first declaration 
about the V\"axj\"o interpretation was partially a consequence of the active advertising of QBism at V\"axj\"o-2001 conference.
My reaction (as many others) was: ``See,  the Copenhagen interpretation finally led to such a 
perverse view on QM as the private agent's perspective on the quantum state.''} However, recently I understood that the situation may be not so bad as one can imagine by reading 
the QBism manifests, such as presented in section \ref{QBINT}. May be this long way to understanding is not only my fault. 
QBists judge too highly the private user interpretation of QM comparing with the problem of concretization of    
the class of such private users. I remember that in 2001 in V\"axj\"o I asked C. Fuchs: ``Suppose that your user, Ivan,  lives in taiga by hunting
and he has never heard about QM. Would Ivan make proper predictions about simplest quantum experiments?''  I do not remember 
the precise answer of C. Fuchs, but it seems it was a long story about his version of FTP (\ref{SIC}), which was considered 
by me as totally irrelevant to my question. Then I asked C. Fuchs the  same question during a few next V\"axj\"o 
conferences and I did not get a satisfactory answer (at least from my viewpoint). Recently, during the V\"axj\"o-2015 conference, where 
QBism was heavily represented by the talks of its founders, I got a new possibility to discuss its foundations with C. Fuchs and 
in the after-conference email correspondence I finally got a clear answer to my old question. Of course, this answer could be found
in the works of Fuchs and Schack, but it was dimmed by the very strong emphasize of the private agent perspective. 

So, if I understood C. Fuchs correctly, the class of agents has to be {\it constrained!} And the basic constraint is given by the Born rule
which is treated as an empirical rule reflecting some basic features of nature, see section \ref{UP} for  discussion.  For a moment,
for us the concrete natural basis of the Born rule is not important. It is important that QBism uses this rule as an information constraint 
to determine  a class of  so to say ``quantum agents'', i.e., those who ``get  tickets to the QBism performance.'' Thus private users of QM are 
those who know the main rule of the game: {\it the probability update for quantum systems has to be done with the aid of the Born rule (or QBist 
version of FTP, see (\ref{SIC})).} It seems reasonable that such agents would produce reasonable predictions. Thus Ivan form taiga is excluded from QBist agents -- finally!

The Born rule constraint is the basic necessary condition for entrance to the QBism club. At the same time it is practically a sufficient conditions, because
other personal characteristics of an agent making predictions about quantum experiments play subsidiary  roles in relation to these predictions. Thus 
in principle one may invent an abstract (conceptual)  QBism-agent who makes her probabilistic predictions about experiment results on the basis 
of the Born rule. In this way QBism comes closely to the recent version of the information interpretation of QM of Zeilinger-Brukner 
proposed by C. Brukner \cite{BR4}. However, while QBistd would, in principle, accept an interpretation a la Brukner, i.e., referred 
to a conceptual agent, they definitely would not like to diminish the role of private agent perspective in QBism. \index{private agent perspective} 
      
Finally, I point to coupling of QBism with recently flowering applications of the probabilistic formalism of quantum mechanics 
outside of physics, see, e.g., \cite{UB_KHR}.   C. Fuchs and R. Schack declared \cite{Fuchs5}:  {\small ``According to QBism, quantum mechanics can be applied to any physical system.
QBism treats all physical systems in the same way, including atoms, beam splitters,
Stern-Gerlach magnets,...''} If so, then why only ``to any physical system''? Why not to any system, biological, cognitive, social, political? 
From my viewpoint, QBism is an excellent interpretation to motivate extension of the domain of applications of the quantum formalism.

However, it seems  that, for a QBist, it is difficult (if possible at all) to accept the possibility of such wider use. The reaction 
of C. Fuchs to my comments in this direction cannot be characterized  as excitement. Of course, this calm reaction might have social 
roots.  By assuming that QBism is a theory about generalized probability update done not only in physics, but, in fact, everywhere, 
QBists would depart even farther from the main stream physics.   

However, it might be that distancing from applications outside of physics has fundamental grounds. In contrast  to the V\"axj\"o interpretation,
in QBism the Born rule is not just a consequence of a very general scheme of the probability update. Its appearance in quantum physics
is a consequence of some fundamental feature of nature, namely, a kind of intrinsic quantum randomness. Thus by extending the domain of applications 
of the quantum formalism to cover, e.g., cognition, one has to assign to cognition a kind of intrinsic quantum(-like) randomness. In principle, one cannot exclude 
that not only quantum physical systems, but even bio-systems are intrinsically random. However, this is a very complicated problem. It seems that QBists (busy with their own problems in physics) simply do not like to be involved in the problem of justification of intrinsic bio-randomness. 

\section{QBism challenge: Born rule or Hilbert space formalism?}
\label{BRHSF}

This section is based on the output of my discussions with A. Plotnitsky on the QBists emphasis of the role of the Born rule 
in QM and their attempts to derive it from the generalized scheme for probability update based on the QBist-FTP (\ref{SIC}). 

In QBism\index{QBism} the Born rule  \index{Born rule} is esteemed very high. It is important to point out that this rule is just a component of the 
complex Hilbert space representation of states of quantum systems (independently of  their interpretation), i.e., it is just 
a part of the {\it mathematical model} serving QM.  If the complex Hilbert space representation
is established (derived or postulated), then Born's rule appears practically automatically:

{\it The wave function cannot be used directly to predict probabilities, because it is a complex vector in a Hilbert space, and you
need a real number to get to probabilities, for which the conjugation is
the easiest way, which is the Born rule.} 

\medskip

Heisenberg already de facto used it for the special case of the probabilities for elections emitting photon in the 
hydrogen atom.  Dirac and  von Neumann, of course, used in connection with the 
 projection postulate.\index{projection postulate} Hence, given the Hilbert space representation of states, 
the Born rule is its trivial consequence. 

In QBism the Born rule is interpreted as an empirical rule. However, the Born rule is not simply 
a feature of nature, but rather, along with the mathematical model itself\footnote{They always work together forming 
a coherent set of constraints upon effective predictions, as the Born
rule is just as meaningless apart from the mathematical model as the use of
the mathematical formalism cannot predict the outcomes of measurements without Born's rule,
see footnote 22 \cite{PL_KHR}}, the rule responds to something that is in
nature and in our interactions with nature, which is, in a way, also part of
nature, insofar as we are part of nature.  It is also important to remark 
 that the Born rule is added to the Hilbert space formalism and not derived from it, but both are
always used together for any prediction.

We also remark that, in contrast to  QBism, the V\"axj\"o approach can be characterized as an attempt 
to derive the Hilbert space representation of probabilities with aid of complex probability amplitudes (wave functions), 
not just the Born rule in, e.g., SIC-POVM form. This assigns a higher value to the wave function Hilbert space representation
than in  QBism. This is a special case of the mathematical modeling approach to study of natural (and mental) phenomena, 
again cf. with V\"axj\"o interpretation.. 

This question can be considered from even more general point:  even if
we have a different mathematical model, somehow without amplitudes (wave functions),
 but still with some Born or Born-like rule (for example, because the SIC-POVM formalism is still
complex-number based), the agent's commitment to this rule will necessarily cover
the mathematical model, too. Arguably, it will be to the
model in the first place, because the rule applies in the framework of the model and
is meaningless without it. For example, one needs to be committed
to the mathematics which defines SIC-POVMs, if only in a trivial sense of
manipulating the formal symbols involved.  On the other hand, we need such a rule to
make predictions in cases like QM. 

In classical mechanics, or relativity, the formalism is descriptive and
predictions follow from this description, so the rule is contained in the formalism.

 May be  QBism current program is to embed the rule in the mathematical model  somehow,
but it is not clear whether this is possible, as the QBism-formalism is clearly not
descriptive. 

That is not only the question of adding probability calculation (the
rule for how amplitudes change), but how to get to probability in
the first place. It was not easy even in classical statistical physics,\index{statistical physics} 
and it took the likes of Boltzmann, Maxwell, Planck, and Gibbs, to
establish these rules.

\section{Is QBism a version of the Copenhagen interpretation?}
\label{MQ7}

By following the talks of C. Fuchs or R. Schack I always had the feeling that their views are very much in the spirit of Copenhagen. 
Bohr and Heisenberg always pointed out that the quantum formalism is not about the ``quantum physical world'', but it is a representation (mathematical) of 
measurements performed on micro-systems. Thus, from their viewpoint QM is about knowledge (especially for Heisenberg). Do Fuchs and his coauthors, Schack, Mermin, Caves,
try to say the same thing by just using the special interpretation of probability, the subjective probability?  I still do not have my own definite opinion about this issue. Therefore 
here I present a long citation of D. Mermin who knows QBism much better than me and who claims that QBism differs crucially from all interpretations in the spirit of Copenhagen, see 
\cite{Mermin}, p. 7-8:   

{\small ``A fundamental difference between QBism and any flavor of Copenhagen, is that QBism
explicitly introduces each user of quantum mechanics into the story, together with the
world external to that user. Since every user is different, dividing the world differently
into external and internal, every application of quantum mechanics to the world must
ultimately refer, if only implicitly, to a particular user. But every version of Copenhagen
takes a view of the world that makes no reference to the particular user who is trying to
make sense of that world.

Fuchs and Schack prefer the term ``agent'' to ``user''.  ``Agent'' serves to emphasize \index{agent} \index{user}
that the user takes actions on her world and experiences the consequences of her actions.
I prefer the term user to emphasize Fuchs' and Schack's equally important point that
science is a user's manual. Its purpose is to help each of us make sense of our private
experience induced in us by the world outside of us.

It is crucial to note from the beginning that ``user'' does not mean a generic body
of users. It means a particular individual person, who is making use of science to bring
coherence to her own private perceptions. I can be a ``user''. You can be a ``user''. But
we are not jointly a user, because my internal personal experience is inaccessible to you
except insofar as I attempt to represent it to you verbally, and vice-versa. Science is
about the interface between the experience of any particular person and the subset of the
world that is external to that particular user.
This is unlike anything in any version of Copenhagen. It is central to the QBist understanding of science.''}

Of course, the reader can find this Mermin's viewpoint on QBism as private agent (user) business does not match completely 
the conclusion from my discussion on the class of users belonging to the ``QBism club'' and constrainted by the knowledge about the Born rule, see 
section \ref{CST}.  This is a consequence of the private user's perspective to interpreting QBism -- both David Mermin and I are good friends of 
the founder of QBism, Christopher Fuchs, and we both got our information about QBism directly from its founder....

\section{Kolmogorov's Interpretation of Probability}
\label{INTK}

Kolmogorov  proposed \cite{K, K1, K2} to interpret probability as follows: ``[. . . ] we may assume
that to an event $A$ which may or may not occur under conditions $\Sigma$, [there] is assigned a real
number $P(A)$ which has the following characteristics: 
\begin{itemize} 

\item (a) one can be practically certain
that if the complex of conditions $\Sigma$ is repeated a large number of times, $N,$ then
if $n$ be the number of occurrences of event $A,$ the ratio $n/N$ will differ very slightly
from $P(A);$

\item (b) if $P(A)$ is very small, one can be practically certain that when conditions
$\Sigma$ are realized only once the event $A$ would not occur.'' 

\end{itemize}

The (a)-part of this interpretation is nothing else than the frequency interpretation of probability, cf. with von Mises
theory and his principle of the statistical stabilization of relative frequencies \cite{[169], [170], [171]}. In the measure-theoretic 
approach this viewpoint on probability is justified by the {\it law of large numbers.} However, for Kolmogorov, approximation 
of probability by frequencies was not the only characteristic feature of probability. The (b)-part (known in foundations of probability 
 as {\it Cournot's principle},\index{Cournot principle} section \ref{COR777}, also plays an important role.
This is the purely weight-type argument: if the weight assigned to an event is very small than one can expect that such an event 
would never happen. We emphasize that  Kolmogorov presented this weight-type argument in its strongest form - ``never happen''.
One may proceed with a weaker form - ``practically never happen'', 
cf.   section \ref{COR777}: weak and strong forms of the Cournot's principle.
 
 \section{Cournot's Principle}
\label{COR777}

We continue to discuss Kolmogorov's interpretation, section \ref{INTK}.
Its (b)-part  is also known as {\it Cournot's principle.}\index{Cournot principle} (In the  presentation of this principle and historical circumstances of its 
appearance we follow the works of G. Shafer, see, e.g., \cite{Shafer}.)
Its first version is due to  J. Bernoulli  (1713) who related mathematical probability to {\it moral certainty/impossibility}\index{moral certainty}\index{moral impossibility}: 

{\small ``Something is morally certain if its
probability is so close to certainty
that the shortfall is imperceptible.''

``Something is morally impossible if
its probability is no more than the
amount by which moral certainty
falls short of complete certainty.''}

By setting the level of moral certainty (or impossibility) Bernoulli connected mathematical probability with the real world:

{\small  ``Because it is only rarely possible to obtain full certainty,
necessity and custom demand that what is merely morally
certain be taken as certain. It would therefore be useful if fixed
limits were set for moral certainty by the authority of the
magistracy|if it were determined, that is to say, whether
99/100 certainty is sufficient or 999/1000 is required. . . ''}

In other words, {\it an event with a small probability will not happen!} \index{small probability}
A.  Cournot (1843) stated that this principle of impossibility is the only way to connect 
mathematical probability with the real world (the name Cournot's
principle was   proposed  by M. Fr\`echet).
A.  Cournot used this principle to justify the frequency interpretation of probability. At that time he already could refer the Bernoulli-Poisson  version 
of the law of large numbers with probability convergence. By using this principle
he concluded  that the event ``the limit of the arithmetic averages would differ from the probability average'' would never happen, since its probability is very small.

This was an important logical step in foundations of probability theory. In 18th-19th centuries, probabilists were not yet able to prove the strong law of large numbers, 
However, they as well as statisticians and physicists badly wanted to apply the frequency interpretation of probability.  The use of Carnot's principle with combination of the weak law of 
large numbers ``solved'' this problem.  

Cournot discussed  not only moral impossibility (very small probability), but also {\it  physical impossibility}
\index{physical impossibility} (infinitely small probability)\index{infinitely small probability}:

{\small ``A physically impossible event is one whose probability is infinitely small.'' }

In spite of the remark,  section \ref{INTK}, that there is a subjective element in the (b)-part of Kolmogorov's interpretation (and Cournot's principle)  - setting the level 
of moral impossibility, those who used this principle treated probability objectively. Subjectivists, as de Finetti, rejected it. 

Cournot's principle principle played an important role in creation of objective interpretation of probability. At the beginning of 20th century it was supported by the majority 
of leading probabilists, e.g., by  Hadamar, L\`evy, Borel, Kolmogorov, Fr\`echet.
Fr\`echet distinguished  between the weak and strong forms of Cournot's principle.
The weak form says an event of small probability seldom happens.  (Fr\`echet, Cramer).
The strong form says an event of small probability will not happen (Cournot, Hadamard, L\`evy, Kolmogorov, Borel).
As well as von Mises,  Fr\`echet and L\`evy agreed that Cournot's principle leads to
an objective concept of probability: Probability is a physical property just like length and weight.

However, in  late 1950s and 1960s Cournot's principle practically disappeared from 
the probability scene. There were a few reasons for this. First of all, Kolmogorov after publication 
of his fundamental book on axiomatics of probability theory did not discuss the interpretational issues
so much (may be not all). As was already pointed out, his model of probability started to be used 
``interpretationally free'' - probability is just a measure. Another reason was strong attacks of subjectivists 
to the objective viewpoint on probability, especially de Finetti's  critique.  Moreover,  as we have seen in Kolmogorov's interpretation,
Cournot's principle is supplementary to the first part of this interpretation which is purely frequentist. And frequentists, as von Mises \cite{[169], [170], [171]}, 
tried to proceed just using the (a)-part of Kolmogorov's interpretation. They neither support Cournot's principle. Finally, after Kolmogorov proved 
the strong version of the law of large numbers, one need not more use this principle as a supplement to the weak version of this law (Bernoulli-Poisson-Khinchin).
At the same time Kolmogorov by himself still explored  Cournot's principle as the second part of his interpretation of probability and it was after he proved 
Theorem 5 \cite{K0}. And moreover the strong law of large numbers was presented in his monograph \cite{K}.\footnote{So, it seems a lot of historicity and psychology was involved in creation of Kolmogorov's theory of probability. I remark that in general his foundational 
monograph makes the impression that it was written ``per one time'', without deep reflections. It really might be the case. Kolmogorov never put his work on creation of the modern
axiomatics of probability theory  to the top of his scientific achievements. In private conversations he prioritized  the contribution to theory of dynamical systems. He also told (again private 
communications) that at that time the idea about the measure-theoretic formalization of probability ``was in the air''. May be the latter is correct, but at the same time if we compare 
his work with works of his main competitors, say von Mises, Borel, Bernstein, Fr\`echet,   we would be surprised by clearness and shortness of Kolmogorov's formalization comparing 
with longly and darkly writings of others.  For example, I heard many times from top Soviet probabilists about Bernstein's axiomatization of probability, preceding Kolmogorov's 
axiomatization. Recently I read Bernstein's book \cite{Bernstein}. It is a real mess... Bernstein tried to make a step towards an abstract presentation of probability theory,
 but his work heavily suffered of longly algebraic  concretizations. Thus after its reading I appreciated Kolmogorov's contribution to foundations of classical probability theory even more. } 

By the 1970s, only Prokhorov (a former student of Kolmogorov) carries Kolmogorov's fame, expressing
the principle in  very special  form: {\it only probabilities close to zero or one are meaningful.}
(If fact, as student of Moscow state university, I heard this interpretational statement, but I treated it as total nonsense. 
For me probability was a measure and nothing more).
 
 \section{Subjective Interpretation of Probability}
 \label{S7}
 
 There are two basic mathematical models of probability, von Mises' frequency model  \cite{[169], [170], [171]} and 
Kolmogorov's measure-theoretic model \cite{K, K1, K2}.  Nowadays  the first one is practically forgotten and the 
second one is widely used in engineering, telecommunications, statistical physics, chemistry, biology, psychology, social science. However, the von Mises approach played the crucial role in development 
of theory of individual random sequences - randomness as unpredictability\index{unpredictability} (impossibility of a successful gambling system), although at the final stage of the mathematical 
formalization of randomness  it was suppressed by two 
other approaches - randomness as complexity\index{complexity} (Kolmogorov) and as typicality\index{typicality} (Martin-L\"of). 

Each scientific theory\index{scientific theory} consists of the two parts, a mathematical model and an interpretation of the mathematical entities.  Both models (of von Mises and Kolmogorov) are 
endowed with the {\it statistical 
interpretation of probability} \index{probability, statistical  interpretation}: if the same complex of experimental conditions (experimental context) is repeated a large number of times than the frequencies
of outcomes approach  the corresponding probabilities. One of the differences  between  the approaches of von Mises and Kolmogorov is that, for the first one the statistical interpretation is the essence of probability, but for the 
second one probability is initially defined as a measure and then one has to prove a {\it theorem}, the strong law of large numbers\index{law of large numbers}, to motivate the frequency interpretation.
We remark that the definition of probability as a measure is heuristically equivalent to assigning weights to elementary events. Thus Kolmogorov's definition of probability is (again heuristically) based
on the procedure of weighting of events.  (Here Kolmogorov generalized Laplace's definition of probability which was based on assigning of {\it equal weights} 
to all elementary events.) However, Kolmogorov did not want to use this weight-like measure-theoretic definition of  probability as its interpretation, with the aid of the law of large 
numbers he moved to the frequency interpretation coinciding with von Mises' ``genuine frequency interpretation'':  

\begin{itemize}
\item (a) {\small ``one can be practically certain
that if the complex of conditions $\Sigma$ is repeated a large number of times, $N,$ then
if $n$ be the number of occurrences of event $A,$ the ratio $n/N$ will differ very slightly
from $P(A);$''}
\end{itemize}
We remark that in the process of creation of his measure-theoretic axiomatics of probability theory Kolmogorov was strongly influenced by von Mises, see \cite{K} and also \cite{INT0}, \cite{INT_KHR}.
Therefore it  might be that the frequency interpretation was also motivated by von Mises theory. The main argument against using the weight-interpretation is that it loses its heuristic attraction 
for continuous sample spaces such as $\Omega=[a, b], \Omega=\mathbf{R}, \Omega= C([a,b])$  (the space of continuous functions on $[a,b]$ which is endowed with, e.g., Winer measure).
Here, for non-discrete probability measure,  the weight of each single elementary event is zero and it is impossible
to jump from zero probability of elementary events to non-zero probability of a non-elementary event.  

In any event  the weight-like viewpoint to probability did not disappear completely from the Kolmogorov interpretation of probability and its trace can be found in the following statement
(known as Cournot's principle, section \ref{COR777}):

\begin{itemize}
\item  (b) {\small ``if $P(A)$ is very small, one can be practically certain that when conditions
$\Sigma$ are realized only once the event $A$ would not occur.''} 
\end{itemize}

This (b)-part of Kolmogorov's interpretation of probability is totally foreign to von Mises' genuine frequency ideology. 

Now we make a point: in a scientific theory the same mathematical model can have a variety of interpretations. In particular, it happened with Kolmogorov's measure theoretic-model 
of probability. Besides of the commonly used statistical interpretation, probability measures can also  be interpreted in the framework of {\it subjective probability theory.}\index{probability, 
subjective interpretation}

This interpretation was used by T. Bayes as the basis of his theory of probability inference, see then  Ramsey
\cite{[12a]},  de Finetti \cite{Finetti},  Savage \cite{[14a]},  Bernardo and Smith
\cite{[15a]} . Here the probability $P(A)$ represents  an {\it agent's personal, subjective degrees of
belief} in non/occurrence of the event $A.$ \index{degree of belief}
In contrast to the statistically interpreted probability which is objective by its nature the subjective probability is by definition not objective, so to say, ``it does not exist in nature'' independently 
of an agent assigning probabilities to events.   This viewpoint on probability is in the direct conflict with von Mises' viewing of probability theory as a theory of natural phenomena, similar to, e.g., 
hydrodynamics. Kolmogorov and the majority of Soviet probabilists  also took the active anti-subjectivist position. Although Kolmogorov treated probability theory as a mathematical theory
(so his viewpoint on probability theory did not coincide with Mises' viewpoint), he also interpreted it as representing objective feature of repeatable phenomena, statistical stability of them. 

At the same time, since the measure-theoretic definition of Kolmogorov probability is heuristically based on the weighting-like procedure for events, it seems that  the subjective interpretation
matches well  the mathematical framework of Kolmogorov probability spaces. Instead of assigning to events  objective weights (as Kolmogorov proposed to do), subjectivists assign to events personal weights, each agent 
assigns to an event $A$ his own degree of belief. This personalization of probability  contradicts not only to the views of von Mises, Kolmogorov and all their followers, but even the basic methodology
of modern science. And, for example, de Finetti understood this well and emphasized this in his exciting and provocative 
 essay \cite{Finetti1}. He started with a citation of the important science methodological statements of Tilgher, see  \cite{Finetti1}, p.169 (in all following citations the italic font was inserted by me):

{\small ``Truth no longer lies in an imaginary equation of the spirit with what is outside it, and which, being outside it, could not possibly touch it and be apprehended; truth is in the very act of 
the thinking thought. The absolute is not outside our knowledge, to be sought in a realm of darkness and mystery; it is in our knowledge itself. {\it Thought is not a mirror in which a 
reality external to us is faithfully reflected}; it is simply a biological function, a means of orientation in life, of preserving and enriching it, of enabling and facilitating action, 
of taking account of reality and dominating it.''}

This viewpoint, thought is just a biological function and not reflection of the objective features of external reality, was shared by de Finetti and used by him to question the conventional
ideology of modern science \cite{Finetti1}, p. 169:      

{\small``For those who share this point of view, which is also mine, but which I could not have framed better than with these incisive sentences of Tilgher's  [..], 
{\it what value can science have?} In what spirit can we approach it? Certainly, we cannot accept determinism; we cannot accept the ``existence'', in that famous alleged realm of 
darkness and mystery, of immutable and necessary  ``laws'' which rule the universe, and we cannot accept it as true simply because, in the light of our logic, it lacks all meaning. 
Naturally, then, science, understood as the discoverer of absolute truths, remains idle for lack of absolute truths. But {\it this doesn't lead to the destruction of science;} 
it only leads to a different conception of science. Nor does it lead to a ``devaluation of science'': there is no common unit of measurement for such disparate conceptions. 
Once the cold marble idol has fallen in pieces, the idol of perfect, eternal and universal science that we can only keep trying to know better, we see in its place, beside us, 
a living creature, the science which our thought freely creates. A living creature: flesh of our flesh, fruit of our torment, companion in our struggle and guide to the conquest.} 

{\small Nature will not appear to it as a monstrous and incorrigibly exact clockwork mechanism where everything that happens is what must happen because it could not but happen, 
and where all is foreseeable if one knows how the mechanism works. To a living science nature will not be dead, but alive; and it will be like a friend about whom one can 
learn in sweet intimacy how to penetrate the soul and spirit, to know the tastes and inclinations, and to understand the character, impulses and abandonments. 
So {\it no science will permit us say: this fact will come about, it will be thus and so because it follows from a certain law, and that law is an absolute truth.} 
Still less will it lead us to conclude skeptically: the absolute truth does not exist, and so this fact might or might not come about, it may go like this or in a totally different way, 
I know nothing about it. What we can say is this: {\it I foresee that such a fact will come about, and that it will happen in such a way, because past experience and its scientific elaboration 
by human thought make this forecast seem reasonable to me.} }

Here the essential difference lies in what the ``why'' applies to: I do not look for 
why THE FACT that I foresee will come about, but why I DO foresee that the fact will come about. It is no longer the facts that need causes; it is our thought that finds it convenient 
to imagine causal relations to explain, connect and foresee the facts. Only thus can science legitimate itself in the face of the obvious objection that our spirit can only think its thoughts, can only 
conceive its conceptions, can only reason its reasoning, and cannot encompass anything outside itself.''
           
This statement contains such charge of energy that even one treating probability objectively cannot reject it without deep analysis. Of course, primarily  de Finetti is right 
that in scientific prediction ``I foresee that such a fact will come about, and that it will happen in such a way, because past experience and its scientific elaboration 
by human thought make this forecast seem reasonable to me.'' We have only our thought and even existence of objective reality is just one of its fruits.
\footnote{I cannot miss the possibility to cite here also Deepak Chopra - controversial New-Age guru (his email from November 3, 2015) :
``Scientific theories are made in consciousness, experiments are designed in consciousness, observations are made in consciousness ,
There is no science without consciousness.
Consciousness created science.
We objectify our human experience in consciousness and called that reality!
There is no reality without consciousness.
Atoms and molecules are words invented by humans for experiences in consciousness.
Using experiences to explain experiences is tautology.
Scientists need more humility and less hubris.''}
For me the essential difference lies in the interpretations of  ``human thought'': either as personalized or as collective. In the above citation from de Finetti, it seems that ``human thought'' has the 
meaning of thought of a kind of the {\it universal agent} doing scientific research. (The same can be said about ``human experience'' in Chopra's statement, see the previous footnote). If we 
take subjective probability as the degree of belief of such a universal agent, then the dispute about objectivity or subjectivity of probability would be resolved peacefully. If de Finetti 
were not assumed the existence of objective reality ruled by natural laws, but just assumed the use of the scientific experience of the mankind, represented as the universal thinking agent, then
von Mises or at least Kolmogorov might agree that such kind of subjective probability has the right for existence. This my playing with the universal agent perspective on the subjective probability
can be compared with Gnedenko's statement, section \ref{GN777}:  {\small ``Subjective probabilities, if necessary, can be made objective.''}

However, (and this the main point) de Finetti strongly supports the personal viewpoint on subjective probability and, hence, ``human thought'' and, ``past experience and its 
scientific elaboration''.\footnote{It is clear why. And now the reader will understand why I did cite Chopra - because Chopra's attempt to escape personalization of human experience
led to global consciousness, a kind of consciousness of universe. De Finetti definitely did not want to be in  the global consciousness club.}          

This {\it personal agent viewpoint} is unsympathetic  for the majority of  scientists, especially those exploring natural sciences. One of the main problems is that subjectivity of probability leads
to {\it subjectivity of cause.} To illustrate this subjectivity of ``causal relations'', de Finetti presented the following provocative example \cite{Finetti1}, pp.179-180:

{\small ``... the essence of the idea of cause escapes me entirely: it only shows itself when I pass from what is already known to predicting the unknown, when the factual data affect 
our state of mind, when from the easy science of hindsight we want to get a rule of action for the future. Suppose it to have been observed that many times, after an eclipse there is a war. 
Why don't I say that the eclipse is a cause of war, and why do superstitious people believe it? And why do we call them superstitious? In saying that the eclipse is not a cause of war 
I mean that, if tomorrow I see an eclipse, the outbreak of war will not therefore seem more likely to me than if an eclipse had not happened. 
One who says that the eclipse is a cause of war would mean that for him, on the contrary, after an eclipse he would see the threat of war as imminent. I call him superstitious 
because his state of mind is different from mine and {\it from that of the society to which I belong,} because it clashes with the conception of the world which is the innermost 
part of my imagination and of the imagination of my century. But if I want to strip away the part of my thought that is my own creation, if I want to distill from my opinions the 
objective part, i.e., the part that is purely logical or purely empirical, I will have to recognize that I have no reason to prefer my state of mind to that of a superstitious 
person except that I actually feel the state of mind which is mine, while that of a superstitious person repels me. The example I gave is an extreme case, 
and it might seem paradoxical. But there are infinitely many others where it would not surprise a contemporary if I said that I do not know how to tell whether or not a causal 
relation exists; there are infinitely many cases that daily give rise to such discussions. I expressed my opinion: {\it the concept of cause is subjective.}'' }

\medskip
   
Here by rejecting the position of  {\small ``eclipse causing believers''} and selecting the position of those   {\small  ``from that of the society to which I belong''} de Finetti might explore the universal agent 
perspective. However, he treated two aforementioned positions as equally acceptable. 
 
The rejection of objectivity of cause by de Finetti can be compared with rejection of causality by von Neumann in his interpretation of QM. 
(Causality is rejected in all versions of the Copenhagen interpretation of QM. However, its probabilistic nature was discussed most clearly in von Neumann book \cite{VN}.)
However, in contrast to de Finetti, by rejecting causality von Neumann did not reject objectivity of probability. He used  the statistical interpretation of probability 
in its genuine von Mises' frequency version.
Von Neumann ``saved'' objectivity of probability in the absence of causality by inveting the concept of {\it irreducible quantum randomness}.\index{irreducible quantum randomness}
Bohr and Pauli also interpreted probability statistically and, for them, it was definitely objective. This objectivity was based on objectivity of outputs of classical measurement devices.  
In contrast to von Neumann, they did not need irreducible quantum randomness.

However, in general de Finetti's denial of objectivity of cause had to be sympathetic for Copenhagenists. Therefore it is surprising that in QM nobody tried to proceed with the subjective probability 
interpretation. Only recently C. Fuchs supported by R. Schack proposed to use in QM subjective probability and personal agent's perspective  (see section  \ref{UP}). This interpretation of QM
is known as Quantum Bayesianism (QBism).
  
For the fathers of QM, both Copenhagenists (as Bohr, Pauli, Dirac, von Neumann) and anti-Copenhagenists (as Einstein, De Broglie),    probability was objective and statistical.
Why? Why not subjective? One of the reasons for this was that all physicists learned probability starting with classical statistical mechanics and the statistical interpretation was firmly 
incorporated in their mind. Some of them were  able to give up even causality, but not statistical nature of probability, 
 However, it seems that the main reason was that de Finetti's views on probability and more generally on scientific theory were {\it too revolutionary} 
even for ``quantum folk''.  The latter still wanted to have solid objective ground - in classical world, the world of macroscopic measurement devices. But de Finetti tried to teach us 
that even in this macro-world neither probability nor cause are objective, they have to be treated subjectively, person dependent. It seems that even Copenhagenists would not accept 
such a position. By following de Finetti consistently, they should reconsider not only physics of microworld, as was done in the process of creation of QM, but even physics of macro-world.
It is interesting that even such a brave guy, C. Fuchs, by exploring the subjective interpretation of probability in quantum physics  was not ready (yet?) to say that now Bayesianism has to 
be extended to classical statistical physics and thermodynamics.\footnote{ 
My personal viewpoint on the subjective  interpretation of probability is sufficiently complicated. As a student of the Department of Mechanics and Mathematics of Lomonosov Moscow 
State University, I was lucky to have a few lectures of Kolmogorov. Then he became too ill to continue, but lectures were given by his former student A. N. Shiryaev; 
in any event  for us ``subjective probability'' was the swearword.  Therefore by working on quantum foundations \cite{INT0},  \cite{KHR_CONT} I always keep the statistical interpretation 
of probability. However, recently by working in applications of quantum probability to cognition, psychology, decision making (Chapter 9) I started to think that the subjective 
interpretation is adequate for modeling of decision making process by an individual agent.  Thus I support Gendenko's statement (section \ref{GN777}): 
{\small ``I think every method has its possibilities and its
limitations.''} }

 \section*{Appendix: Symmetric Informationally Complete Quantum Measurements}
\label{SIC7}

We consider  one special class of  atomic instruments with  quantum observables given by  
{\it symmetric informationally complete} POVMs, SIC-POVMs. Here informational completeness means that the probabilities of observing 
the various outcomes (given by Born's rule) entirely determine any quantum state $\rho$ being measured. 
This requires $d^2$ linearly independent operators for the state space of the dimension $d.$  

The simplest definition  is that a SIC-POVM  is determined by   a system of $d^2$ normalized vectors $(\phi_i)$ (they are not orthogonal) such that 
\begin{equation}
\label{humg1}
\langle \phi_i \vert \phi_j\rangle^2 =  \frac{1}{d+1}, i \not=j.   
\end{equation}
The elements of the corresponding SIC-POVM $(E_i)$ are subnormalized projectors 
$E_i = \frac{1}{d}  \Pi_i,$
where $ \Pi_i $ is the orthogonal projector on $\phi_i.$ 
The elements of SIC-POVM $E_i$ determine the corresponding quantum operations (atomic instruments).

The characteristic property of SIC-POVMs, symmetry,  is that  
 the inner product in the space of operators (or $d\times d$ matrices) given by the trace is constant, 
i.e., 
$$\rm{Tr} E_i E_j = \rm{const}= \frac{1}{d^2(d+1)} , i \not=j.$$
By using this equality it is easy to obtain the following representation of an arbitrary density operator $\rho:$ 
\begin{equation}
\label{SIC_FTP1}
\rho=  \sum_i \Big((d+1) p(i)  - \frac{1}{d}\Big)  \Pi_i
\end{equation}
where $p(i)= \rm{Tr} E_i \rho$ is the probability to obtain 
the result $i$ for a measurement presented by the SIC-POVM $(E_i).$

This SIC-POVM based representation of a density operator $\rho$ plays the crucial role 
in Quantum Bayesianism (QBism), cf. with the QBist version of quantum generalization of  FTP (see section \ref{UP}).

\end{document}